\documentclass[12pt]{iopart}

\usepackage{iopams}
\usepackage{graphicx}

\begin{document}

\title{Super persistent transient in a master-slave configuration with 
Colpitts oscillators}

\author{R. C. Bonetti$^1$, S. L. T. de Souza$^2$, A. M. Batista$^3$,
J. D. Szezech Jr.$^3$, I. L. Caldas$^4$, R. L. Viana$^5$, S. R. Lopes$^5$,
M. S. Baptista$^6$}
\address{$^1$ P\'os-Gradua\c c\~ao em Ci\^encias, Universidade Estadual de 
Ponta Grossa, 84030-900, Ponta Grossa, Paran\'a, Brazil}
\address{$^2$ Departamento de F\'isica e Matem\'atica, Universidade Federal de 
S\~ao Jo\~ao del Rei, 36420-000, Ouro Branco, Minas Gerais, Brazil}
\address{$^3$Departamento de Matem\'atica e Estat\'istica, Universidade 
Estadual de Ponta Grossa, 84030-900, Ponta Grossa, Paran\'a, Brazil}
\address{$^4$Instituto de F\'isica, Universidade de S\~ao Paulo, Caixa Postal 
66316, 05315-970, S\~ao Paulo, SP, Brazil}
\address{$^5$Departamento de F\'isica, Universidade Federal do Paran\'a,
81531-990, Curitiba, Paran\'a, Brazil}
\address{$^6$Institute for Complex Systems and Mathematical Biology, 
SUPA, University of Aberdeen, AB24 3UE Aberdeen, United Kingdom}
\ead{$^3$antoniomarcosbatista@gmail.com}

\begin{abstract}
Master-slave systems have been intensively investigated to model the
application of chaos to communications. We considered Colpitts oscillators 
coupled according to a master-slave configuration to study chaos 
synchronisation. We revealed the existence of super persistent transients in 
this coupled system. Moreover, we showed that an additive noise added to the 
slave system may suppress chaos synchronisation. When synchronisation is not 
suppressed, the noise induces longer transients.
\end{abstract}

\hspace{4pc}{\small {\it Keywords}: Colpitts, synchronisation, master-slave}

\maketitle

\section{Introduction}

Coupled chaotic systems can synchronise their trajectories \cite{pecora90}. 
The synchronisation of coupled chaotic systems has important applications in 
many fields, such as: biological systems \cite{batista12}, secure 
communications \cite{yang13}, chemical reactions \cite{wang10}, etc.

Recent works have shown that chaos synchronisation of a master-slave 
configuration is relevant to communication systems. The slave system is driven 
by a signal derived from the master \cite{suy}. Here we focus the study in two 
chaotic coupled Colpitts oscillators \cite{feo}. We choose this circuit due to 
the fact that it can be useful in applications to communication systems, as 
well as, it exhibits a rich dynamical behaviour for certain parameter values
\cite{feo2003}. In the Colpitts oscillator the operation frequency can vary 
from few Hertz up to the microwave frequency range, a characteristic that 
enables the use of this circuit to transmit information in channels with 
difference frequency bandwidths.

There has been a great interest in the study of synchronisation and control of 
Colpitts oscillators. Control schemes were used to suppress chaos. Li and 
collaborators used a controller to drive chaotic Colpitts to a desired state 
\cite{li}, that is, the stabilisation of the chaotic motion to a steady state. 
In another work, the circuit was controlled by using a non linear feedback 
\cite{effa}. It has been found synchronisation between chaotic Colpitts in 
identical and mismatched cases \cite{baziliauskas,machuca}. Furthermore, there 
are works about synchronisation of Colpitts oscillators that operate in ultra 
high frequency range \cite{kengne}.

In this article, we study two coupled Colpitts oscillators in a master-slave 
configuration and focus our attention to chaos synchronisation. Our main 
objective is to verify the existence of super persistent chaotic transients 
\cite{grebogi,andrade}, and the effect of noise on the synchronous behaviour 
\cite{souza01,souza07}.

This article is organized as follows: in section 2 we present the coupled 
Colpitts oscillators. In section 3 we study the onset of synchronisation and 
show the existence of super persistent transients. In section 4 we describe 
the effect of noise on the system. The last section contains our conclusions. 

\section{Colpitts oscillator}

\begin{figure}
\begin{center}
\includegraphics[width=20pc,height=15pc]{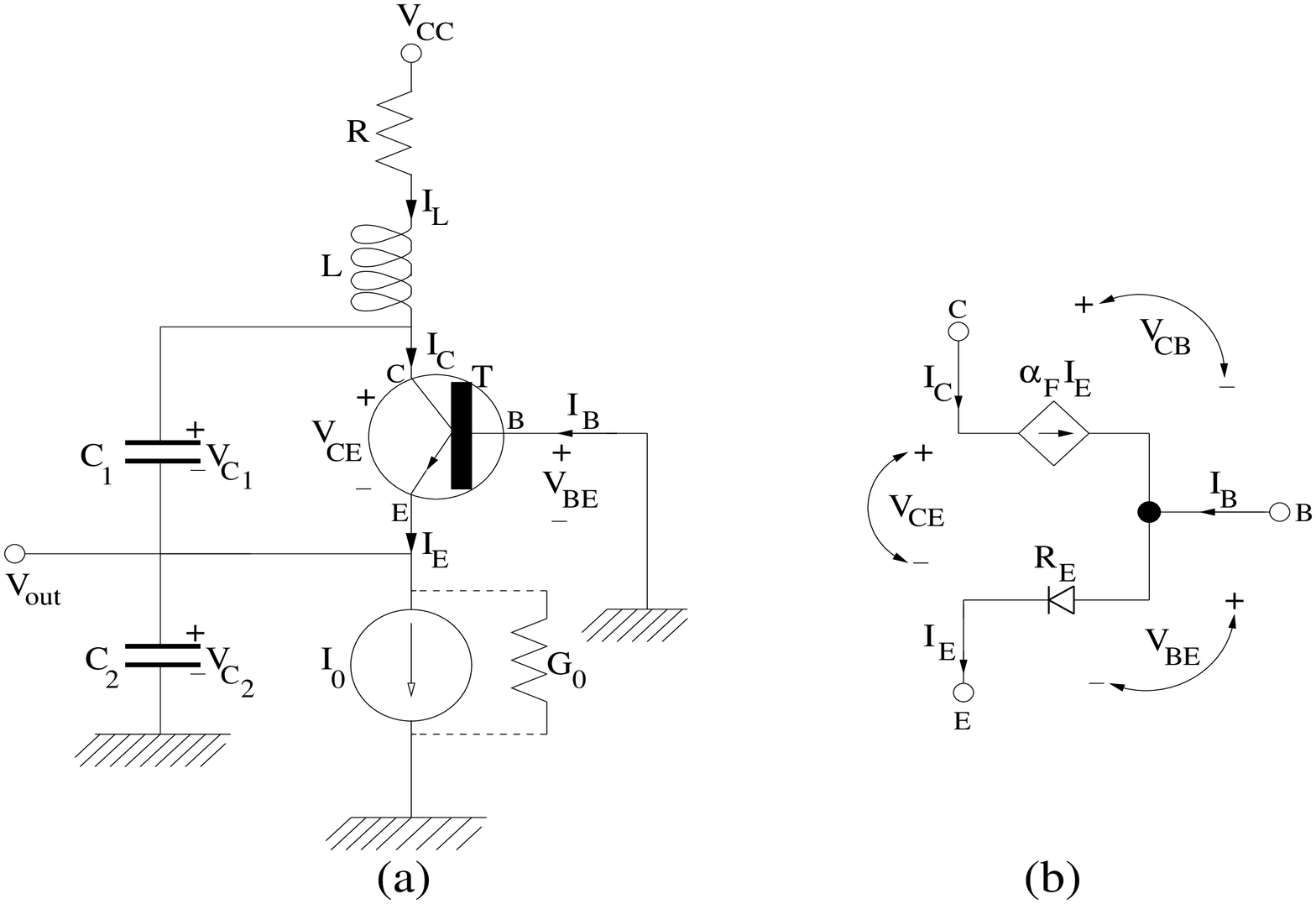}
\caption{The circuit diagram of a Colpitts oscillator. (a) Circuit 
configuration and (b) the bipolar junction transistor (BJT).}
\label{fgr1} 
\end{center}
\end{figure}

The Colpitts oscillator is a type of resonant circuit with a transistor for
feedback. This oscillator has been used in electronic devices and communication
systems, due to the fact that it can exhibit chaos \cite{kennedy}. Fig. 
\ref{fgr1}(a) exhibits the circuit configuration containing a bipolar junction 
transistor (BJT) according to Fig. \ref{fgr1}(b).
The state equations are given by
\begin{eqnarray}
C_{1}{\frac{dV_{C_1}}{dt'}} & = & -\alpha_{F}f(-V_{C_2})+I_{L},\nonumber\\
C_{2}{\frac{dV_{C_2}}{dt'}} & = & (1-\alpha_{F})f(-V_{C_2})-G_{0}V_{C_{2}}+I_{L}
-I_{0},\nonumber\\
L{\frac{d{I_{L}}}{dt'}} & = & -V_{C_1}-V_{C_2}-RI_{L}+V_{CC},
\label{eqestcolp}
\end{eqnarray}
where the voltages $V_{C_{1}}$ and $V_{C_{2}}$ are associated with the capacitors
$C_1$ and $C_2$, respectively, $V_{cc}$ is the voltage supply, $I_{L}$ is the 
current through the inductor $L$ and $t$ is the time. There is a current 
generator $I_0$ to maintain constant biasing emitter current. The function 
$f()$ is the driving-point characteristic of the non linear resistor $R_E$ and 
it can be expressed as $I_{E}=f(V_{C_{2}})=f(-V_{BE})$, where $\alpha_{F}$ is the 
common-base forward short-circuit gain.

\begin{figure}
\begin{center}
\includegraphics[width=20pc,height=14pc]{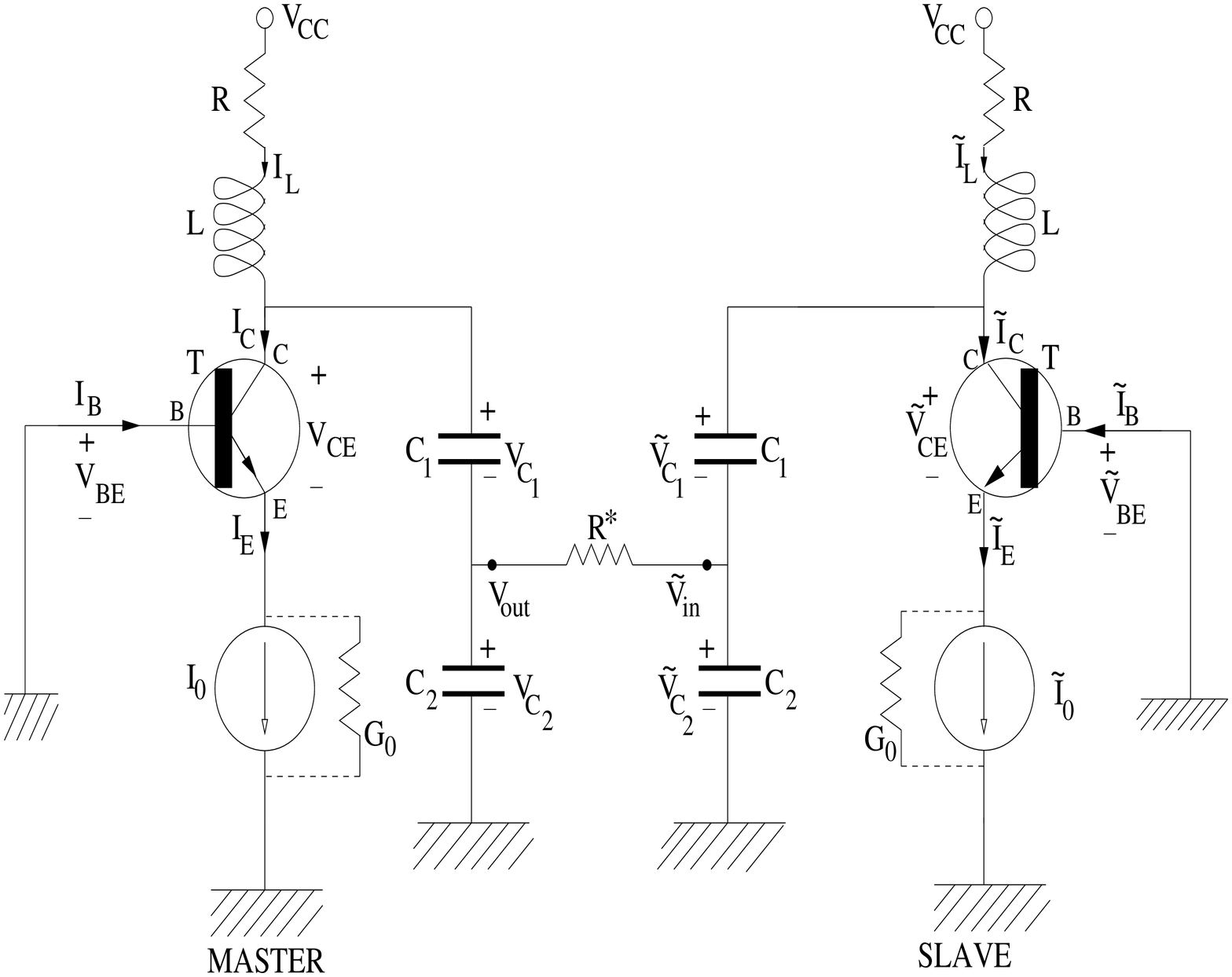}
\caption{The circuit diagram of coupled Colpitts with a master-slave
configuration.}
\label{fgr2} 
\end{center}
\end{figure}

We consider two Colpitts oscillators in a master-slave configuration, in
accordance with the scheme sho\-wed by Fig. \ref{fgr2}. If this coupled 
system would be used for communication purposes, the transmitter would be the
master and the receiver the slave. We also consider noise in the channel.

Introducing the following dimensionless variables
\begin{eqnarray}
x_{1}&=&\frac{V_{C_1}-\overline{V}_{C_1}}{V_{T}},\nonumber \\
x_{2}&=&\frac{V_{C_2}-\overline{V}_{C_2}}{V_{T}},\nonumber\\
x_{3}&=&\frac{I_{L}-\overline{I}_{L}}{I_{0}},
\label{normm}
\end{eqnarray}
where 
\begin{eqnarray}
\overline{V}_{C_1}&&=V_{CC}-\alpha_{F}Rf(-\overline{V}_{C_2})-\overline{V}_{C_2},\\
\overline{V}_{C_2}&=&\frac{1}{G_{0}}\bigg[n(-\overline{V}_{C_2})-I_{0}\bigg]\\
\overline{I}_{L}&&=-\alpha_{F}f(-\overline{V}_{C_2}),
\label{eqpontoeq}
\end{eqnarray}
we obtain the equations of an unidirectional master-slave configuration 
\begin{eqnarray}
\frac{dx_{1}}{dt}&=&{\frac{g}{Q(1-k)}}[-\alpha_{F}n(x_2)+x_3],\nonumber\\
\frac{dx_{2}}{dt}&=&{\frac{g}{Qk}}[(1-\alpha_{F})n(x_2)+x_3],\nonumber\\
\frac{dx_{3}}{dt}&=&-{\frac{Qk(1-k)}{g}}[x_{1}+x_{2}]-{\frac{1}{Q}{x_3}},
\nonumber\\
\frac{dy_{1}}{dt}&=&{\frac{g}{Q(1-k)}}[-\alpha_{F}n(y_2)+y_3],\nonumber\\
\frac{dy_{2}}{dt}&=&{\frac{g}{Qk}}[(1-\alpha_{F})n(y_2)+y_3]+\varepsilon [x_{2}
-y_{2}],\label{eqms}\\
\frac{dy_{3}}{dt}&=&-{\frac{Qk(1-k)}{g}}[y_{1}+y_{2}]-{\frac{1}{Q}{y_3}},
\nonumber\\
t&=&t'\omega_0, \\
\omega_0&=&\frac{1}{\sqrt{L\frac{C_1C_2}{C_1+C_2}}},
\label{eqcolpacolp}
\end{eqnarray}
where $x_1$,$x_2$, and $x_3$ belong to the master circuit, $y_1$,$y_2$ and 
$y_3$ belong to the slave circuit, $\omega_0$ is the resonant frequency of the 
tank circuit due to $L$, $C_1$ and $C_2$. The time derivative of $y_2$ 
containing the coupling term depends on variables of both circuits.
The non linear terms are given by 
$n(x)=e^{-x}-1$, and $n(y)=e^{-y}-1$. The dimensionless parameters are 
\begin{eqnarray}
Q&=&\frac{\omega_{0}L}{R}, \\
k&=&\frac{C_{2}}{{C_{1}+C_{2}}}, \\
g&=&\frac{I_{0}L}{V_{T}R(C_{1}+C_{2})},
\end{eqnarray}
and $\varepsilon$ is the coupling strength, that is defined as 
$\varepsilon=\sqrt{L/C_1}R^{*-1}$. We use $k=0.5$, $\alpha_{F}=0.996$, 
$R=80\Omega$, $C_{1}=C_{2}=1\mu F$, $L=18.2 \mu H$, $V_{T}=27 mV$, and $Q=1.77$. 
With these value set and $g=2.863$ the Colpitts oscillator presents chaotic 
behaviour. The system presents a broad range of different dynamical regimes as 
the parameter $g$ is varied, such as chaotic behaviour, periodic solutions, 
hopf bifurcation, and coexistence of solutions. In this work, in order to study 
chaos synchronisation, we consider a small interval of the values of $g$ such 
that the system only presents chaotic behaviour.

\section{Chaos synchronisation}

Chaotic systems have applications in secure and spread spectrum communications.
Previous works presented applications of chaos synchronisation in wireless 
communications \cite{sano} and multiplexing mixed chaotic signals generated by 
different electronic oscillators \cite{shi}. Chaos synchronisation occurs when 
the state variables of both circuits are equal. Such condition is achieved 
after a transient time that depends on the stability of the synchronisation 
manifold.

\begin{figure}
\begin{center}
\includegraphics[width=18pc,height=20pc]{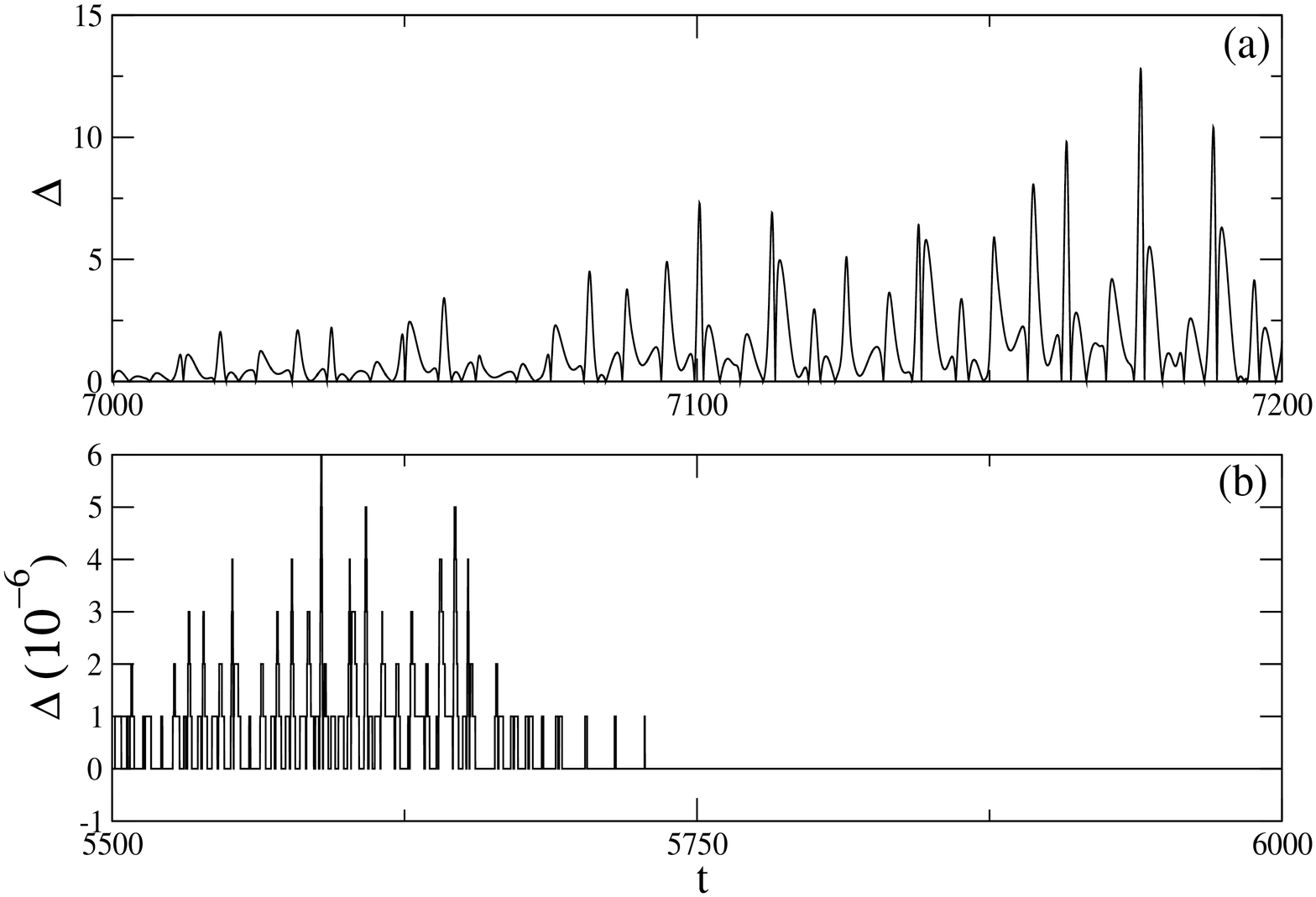}
\caption{Time evolution of the synchronisation error considering (a)
$\varepsilon=0.05$ and (b) $\varepsilon=0.089$.}
\label{fgr3} 
\end{center}
\end{figure}

A numerical diagnosis to state about the synchronous state is provided by the 
synchronisation error 
\begin{equation}
\Delta=|x_1-y_1|.
\end{equation}
Figure \ref{fgr3} shows the time evolution of $\Delta$ for two situations:
(a) when there is no synchronisation between the Colpitts oscillators and (b) 
when there is chaos synchronisation. When the oscillators are completely 
synchronised we have $\Delta=0$.

\begin{figure}
\begin{center}
\includegraphics[width=17pc,height=14pc]{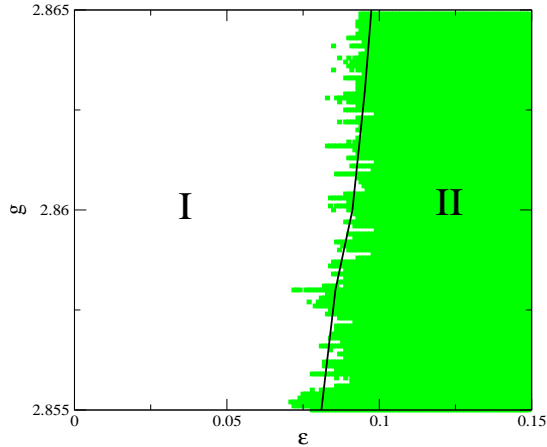}
\caption{(Colour online) Synchronised domains (green region) in parameter plane 
$g\times\varepsilon$. The black line separates the regions for which the 
synchronisation manifold is unstable (I) and stable (II). Stability is 
measured by the Lyapunov exponents of the synchronisation manifold.}
\label{fgr4} 
\end{center}
\end{figure}

We investigate the dependence of the synchronisation error on the control
parameter $g$ and the coupling strength $\varepsilon$. Using the time averaged 
error
\begin{equation}
\bar\Delta=\frac{1}{t_2-t_1}\sum_{t_1}^{t_2}\Delta(t),
\end{equation} 
where $t_2-t_1$ is the time window for measurements, chaos synchronisation is
stated when $\bar\Delta<10^{-4}$. We consider $t_1=5000$ and $t_2=10000$ 
but similar results were obtained for $t_1=18000$ and $t_2=20000$.
Figure \ref{fgr4} shows a parameter space indicating by colours regions of 
parameters that lead to no synchronisation (white) and regions of parameters 
that lead to synchronisation (green), representing parameters for which 
$\bar\Delta<10^{-4}$.

We calculate the spectrum of Lyapunov exponents of the synchronisation manifold
and its transversal directions in order to verify the local stability of the 
synchronisation manifold. We obtain the spectrum considering the same initial 
conditions for both circuits: $x_1=y_1=0.02$, $x_2=y_2=10^{-4}$ and 
$x_3=y_3=10^{-4}$. We are interested in the two largest Lyapunov exponents. When
the maximal exponent is positive and the second largest is negative the system 
presents chaos synchronisation \cite{gade}. The synchronisation manifold is 
locally stable, since the negative exponent measures how perturbation 
propagate along the direction transversal to the synchronisation manifold. 
Consequently, the circuits can synchronise. The black line, shown in Fig. 
\ref{fgr4}, separates two regions. In region I the coupled oscillators do not 
synchronise, while in region II synchronisation occurs.

We can see in Fig. \ref{fgr4} that the boundary between the two regions has an 
irregular pattern. This suggests the existence of an entangled basin 
boundary for the synchronous attractor \cite{camargo}. Consequently, the time 
to reach the synchronous state will strongly depend on the initial conditions, 
some of them responsible for very long transients. The transient time is 
denoted by $\tau$, and its average value is $\tau_M$.

\begin{figure}
\begin{center}
\includegraphics[width=16pc,height=14pc]{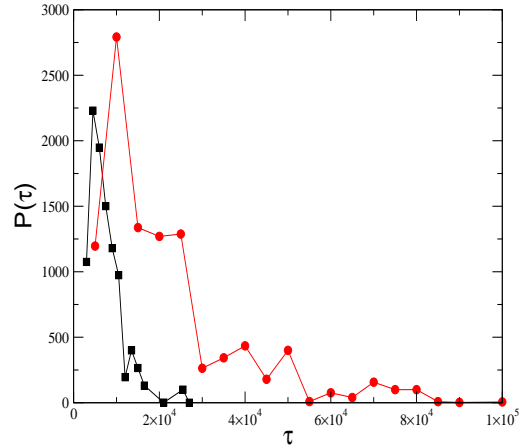}
\caption{(Colour online) Histogram of the transient time intervals for 
a total of $10^4$ different initial conditions $x_2$ and $y_2$ in the interval
$[0,0.001]$, where we consider $g=2.863$, $\varepsilon=0.086$ (red circles)
and $\varepsilon=0.090$ (black squares).}
\label{fgr5} 
\end{center}
\end{figure}

The histogram of the transient time for an ensemble of initial conditions is 
shown in Fig. \ref{fgr5}, where the red circles correspond to 
$\varepsilon=0.090$ and the black squares to $\varepsilon=0.086$. The 
statistical distribution of the transient sizes was obtained through $10^4$ 
different initial conditions of $x_2$ and $y_2$, showing that small transients 
are more common when the value of the coupling strength increases. Therefore, 
for some values of $g$ and $\varepsilon$, around the region of the boundary
in the green area of Fig. \ref{fgr4}, the circuits may present large transient 
depending on the initial conditions. 

\begin{figure}
\begin{center}
\includegraphics[width=16pc,height=14pc]{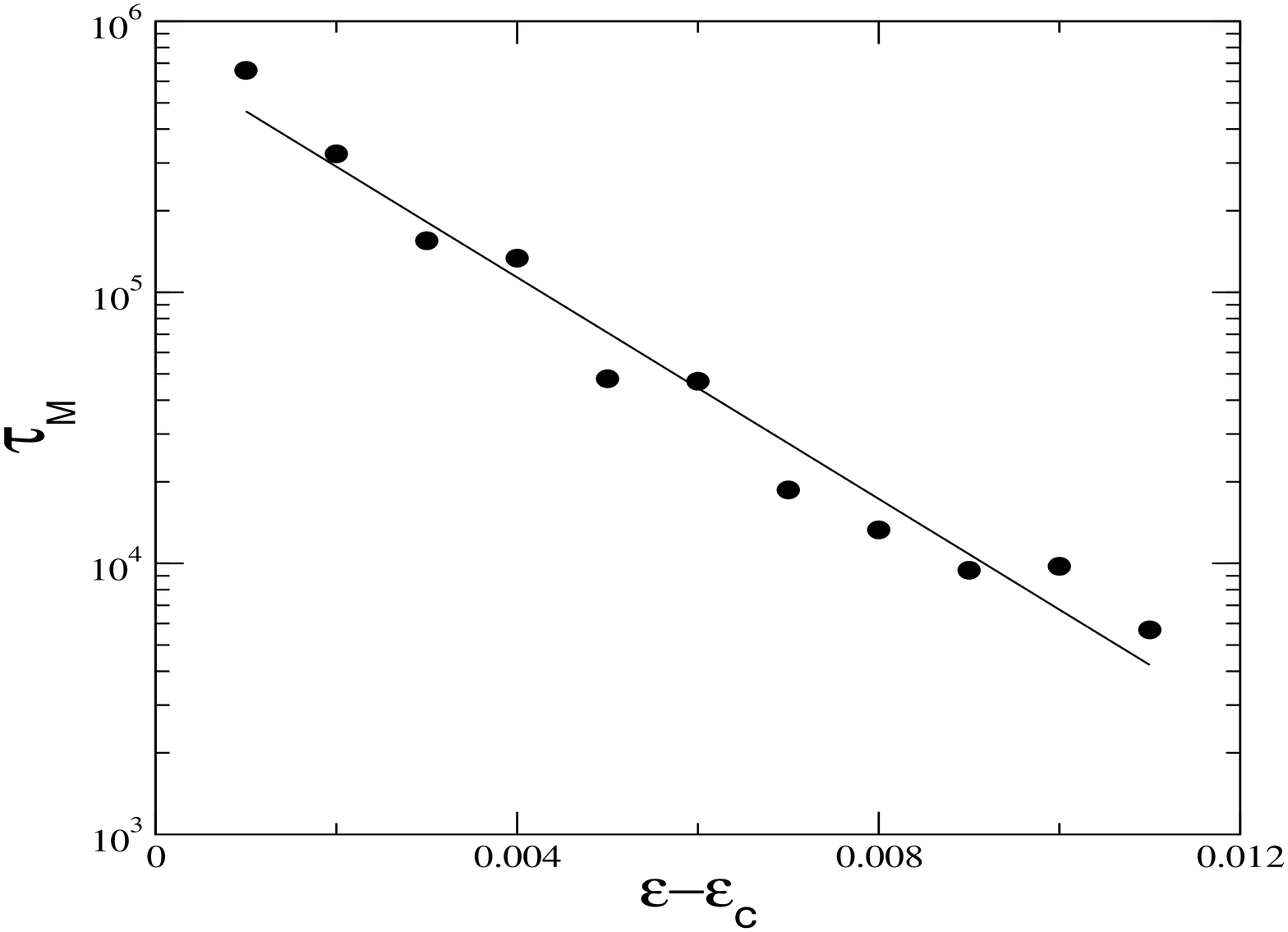}
\caption{Scaling of super persistent transient varying the coupling strength
for $g=2.863$ and $\varepsilon_c=0.078$. Each point represents the average 
over 100 different initial conditions of $x_2$ and $y_2$. The solid line is an
exponential fitting with exponent $-470.31$. The value of the $\varepsilon_c$
is obtained when the synchronisation manifold becomes stable.}
\label{fgr6} 
\end{center}
\end{figure}

Here we recall that there is a distinct class of chaotic transients that is 
referred to as super persistent \cite{grebogi,andrade}. This transient times is
characterized by the following scaling law for their average transient
\begin{equation}
\tau_M\sim \exp [\beta(p-p_c)^{-\gamma}],
\end{equation}
where $\beta$ and $\gamma$ are positive constants, $p_c$ is a critical parameter
value, and the transient occurs for $p>p_c$. The least-squares fit in 
Fig. \ref{fgr6} exhibits an exponential distribution for $\tau_M$ and
$\varepsilon-\varepsilon_c$, which indicates a super persistent 
transient with exponent $\gamma=1$.

\section{The effect of noise}

We analyse chaos synchronisation under realistic conditions, due to noise in 
experiments \cite{hong}. We add a stochastic perturbation in the variable 
$y_2(t)$ of equation (\ref{eqms}) 
\begin{equation}
\frac{dy_{2}}{dt}={\frac{g}{Qk}}[(1-\alpha_{F})n(y_2)+y_3]+\varepsilon [x_{2}
-y_{2}]+Ar(t),
\end{equation}
where $A$ is the level of the stochastic perturbation and $r(t)$ is a 
pseudo-random variable.  We consider a random number generator that returns a 
normally distributed deviate with zero mean and unit variance. In other words, 
the noise is a white Gaussian noise (AWGN) \cite{shi08}.

\begin{figure}
\begin{center}
\includegraphics[width=15pc,height=14pc]{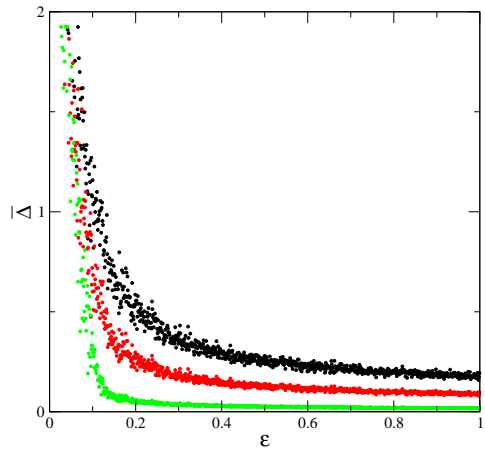}
\caption{(Colour online) Synchronisation error versus coupling strength for 
$g=2.863$, $A=0.1$ (green circles), $A=0.5$ (red circles) and $A=1.0$ (black 
circles). These three cases have exactly the same initial conditions.}
\label{fgr7} 
\end{center}
\end{figure}

To understand the effect of noise in the synchronisation conditions, we measure
the synchronisation error for different values of noise level and coupling
strength. In Fig. \ref{fgr7}, we plot the time averaged synchronisation error 
versus the coupling strength for three values of the noise level. When the 
noise level is small (green circles) the value of the time averaged error 
decreases sharply when the coupling str\-ength increases. For larger values of 
$A$ (red and black circles) the time averaged error decays less abruptly as a 
function of $\varepsilon$. For all situations, $\bar\Delta>0$, showing that 
synchronisation is suppressed.

\begin{figure}
\begin{center}
\includegraphics[width=17pc,height=14pc]{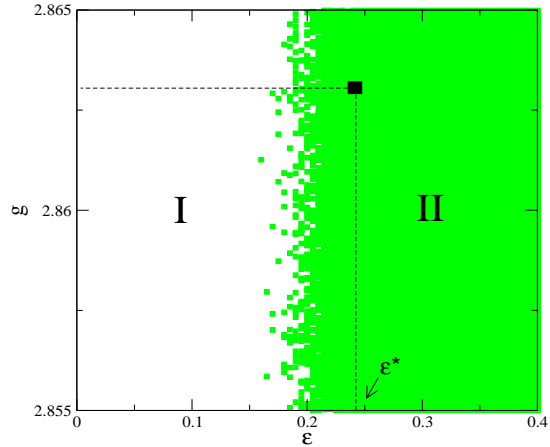}
\caption{(Colour online) Synchronised domains (green region) considering the
same parameters according to Fig. \ref{fgr4}, and a small noise $A=3\times 
10^{-5}$.}
\label{fgr8} 
\end{center}
\end{figure}

\begin{figure}
\begin{center}
\includegraphics[width=18pc,height=15pc]{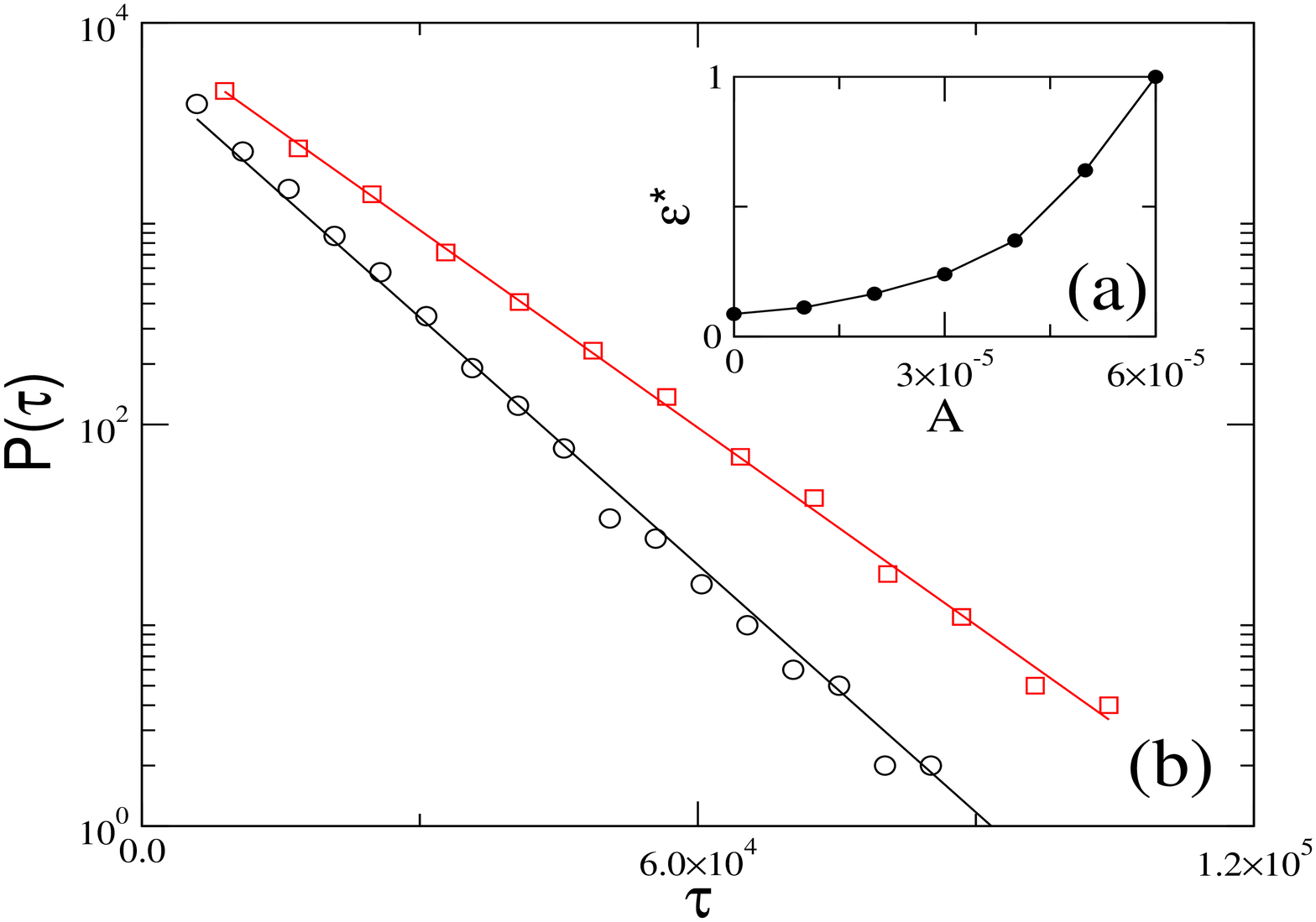}
\caption{(Colour online) (a) $\varepsilon^*$ versus $A$, where the black 
circles correspond to $\tau_M\approx 10^4$. (b) Histograms for transient time 
considering  $g=2.863$, $\varepsilon=1$, $A=5\times 10^{-5}$ (black circles), 
and $A=6\times 10^{-5}$ 
(red squares).}
\label{fgr9} 
\end{center}
\end{figure}

Synchronisation is affected by noise. In Fig. \ref{fgr8} we consider the same 
parameters used in Fig. \ref{fgr4} but add noise with level $A=3\times 10^{-5}$.
Comparing Fig. \ref{fgr8} (with noise) with Fig. \ref{fgr4} (without noise)
it is possible to observe that due to the effect of the noise, a larger value  
of $\varepsilon$ is necessary to make the systems synchronise.
Similar to Fig. \ref{fgr4}, the structure of the boundary between the 
synchronous and the non-synchronous region is a consequence of the existence 
of entangled basin boundary that produces super persistent transients. We 
obtain the synchronised region by verifying if $\bar{\Delta}<10^{-4}$, for 
$t_1=1000$ and $t_2=20000$. We analyse the noise effect on the synchronised
region fixing the value of $g$ and varying the noise amplitude $A$ to
obtain the $\varepsilon^*$ in that the average transient $\tau_M$ is
approximately $10^4$ (Fig. \ref{fgr9}a). Therefore, the larger the noise
amplitude is, the larger the coupling strength must be to synchronise.

The phenomenon of super persistent transient is effected by the noise, that is, 
a small noise increases the transient times. Fig. \ref{fgr9}(b) shows 
histograms for $g=2.863$, $\varepsilon=1$, $A=5\times 10^{-5}$ (black circles), 
and $A=6\times 10^{-5}$ (red squares), in order to show that the noise induces 
longer transients. When $A\leq 4\times 10^{-5}$ the transients have values 
around $10^3$. Moreover, for $A\geq 7\times 10^{-5}$ the noise may suppress 
chaos synchronisation. It is worth to comment that near the critical parameter 
$\varepsilon_c$, the average transient $\tau_M$ quickly increases when the 
noise amplitude grows, according to an exponential relationship 
$\tau_M\sim \phi \exp (\varphi A)$, where $\phi$ and $\varphi$ are positive 
constants.

\section{Conclusions}

In this paper, we studied some aspects of chaos synchronisation displayed
by two coupled Colpitts with master-slave configuration. We obtained a set of
parameters which may lead the coupled circuits to a synchronised or non 
synchronised state. We verified the existence of super persistent transients. 
This transient is mainly situated in the border of the synchronised domain 
in the parameter space $g$ versus $\varepsilon$, where $g$ is the loop gain
of the oscillator, and $\varepsilon$ is the strength coupling.

The effect of noise on the coupled circuits were considered. Noise acts on the 
system in a way that synchronisation can only be achieved for higher coupling 
strength. Moreover, the transients become longer.

Our results enable us to predict a set of parameters of the coupled Colpitts 
oscillators to observe super persistent transients that can be used in 
laboratory experiments. The described persistent transients are similar to
those observed in dissipative systems, and should be related to the chaotic
saddle of the coupled systems \cite{suso02,suso03}.

In future works, we plan to study this coupled system considering electronic 
simulations \cite{kenfack}, and also to get experimental results through an 
electronic circuit.

\ack
This work was possible by financial support from following Brazilian 
government agencies: CNPq, CAPES, Funda\c c\~ao Arauc\'aria (Paran\'a) and
FAPESP. M. S. Baptista acknowledges EPSRC-EP/I032606/1.

\end{document}